\documentclass[fleqn,usenatbib]{mnras}

\usepackage{amsmath}
\usepackage{gensymb}
\usepackage{graphicx}
\graphicspath{{figs/}}
 \usepackage{float}
\usepackage{verbatim}
\usepackage{amssymb}
\usepackage{lineno}
\usepackage{wasysym}
\usepackage{amssymb}
\usepackage{xcolor}
\usepackage{soul}
\usepackage{amstext}
\usepackage{array}
\usepackage{subfig}
\usepackage{enumitem}
\setlist[enumerate,1]{label={\arabic*.}}
\setlist[enumerate,2]{label={(\alph*)}}

\usepackage[T1]{fontenc} 
\usepackage{lmodern} 
\rmfamily 
\DeclareFontShape{T1}{lmr}{b}{sc}{<->ssub*cmr/bx/sc}{}
\DeclareFontShape{T1}{lmr}{bx}{sc}{<->ssub*cmr/bx/sc}{}

\newcommand{\mgii}{\ion{Mg}{ii}}

\newcommand{\sgas}{SGAS\,J1226+2152}

\newcommand{\zabs}{0.77138}

\newcommand{\msun}{M$_{\astrosun}$}

\title[Orientation effects on MgII absorption]{Orientation effects on 
cool gas
  absorption from gravitational-arc tomography of a $z=0.77$ disc galaxy}

\author[]{ 
A. Fernandez-Figueroa,$^{1}$
S. Lopez,$^{1}$\thanks{E-mail: slopez@das.uchile.cl}
N. Tejos,$^{2}$
T.~A.~M. Berg,$^{1,3}$
C. Ledoux,$^{3}$
\newauthor
P. Noterdaeme,$^{4,5}$
A. Afruni,$^{1}$
L. F. Barrientos,$^{6}$
J. Gonzalez-Lopez,$^{7,8}$
\newauthor
M. Hamel,$^{1}$ 
E. J. Johnston,$^{8}$
A. Katsianis,$^{9}$
K. Sharon$^{10}$
and
M. Solimano$^{8}$
\\
$^{1}$ Departamento de Astronom\'ia, Universidad de Chile,
    Casilla 36-D, Santiago, Chile\\
$^{2}$ Instituto de F\'isica, Pontificia Universidad Cat\'olica de
    Valpara\'iso, Casilla 4059, Valpara\'iso, Chile\\
$^{3}$ European Southern Observatory, Alonso de C\'ordova 3107, Vitacura, Casilla 19001, Santiago, Chile\\
$^{4}$ Franco-Chilean Laboratory for Astronomy, IRL 3386, CNRS and Universidad de Chile, Santiago, Chile\\
$^{5}$ Institut d'Astrophysique de Paris, CNRS-SU, UMR 7095, 98bis bd Arago, 75014 Paris, France\\
$^{6}$ Instituto de Astrofı\'isica, Pontificia Universidad Cat\'olica de
    Chile, Av. Vicu\~na Mackenna 4860, 7820436 Macul, Santiago, Chile \\
$^{7}$    Las Campanas Observatory, Carnegie Institution of Washington, Casilla 601, La Serena, Chile\\\
$^{8}$ N\'ucleo de Astronom\'ia de la Facultad de Ingenier\'ia y Ciencias,
    Universidad Diego Portales, Av. Ej\'ercito Libertador 441, Santiago, Chile \\
$^{9}$ Tsung-Dao Lee Institute, Shanghai Jiao Tong University, Shanghai 200240, China\\
$^{10}$ Department of Astronomy, University of Michigan, 1085 South University
    Avenue, Ann Arbor, MI 48109, USA\\ 
}

\begin{document}
\label{firstpage}
\pagerange{\pageref{firstpage}--\pageref{lastpage}}
\maketitle

\begin{abstract}
We use spatially-resolved spectroscopy of a distant giant gravitational arc to
test orientation effects on \mgii\ absorption equivalent width (EW) and covering
fraction ($\langle\kappa\rangle$) in the circumgalactic medium of a foreground 
star-forming galaxy (G1) at
$z\sim0.77$. Forty-two spatially-binned arc positions uniformly sample impact
parameters ($D$) to G1 between $10$ and $30$ kpc and azimuthal angles $\alpha$ between
$30^{\circ}$ and $90^{\circ}$ (minor axis).  We find an EW-$D$ anti-correlation, 
akin to that observed  statistically in quasar absorber studies, and {an apparent} correlation of
both EW and $\langle\kappa\rangle$ with $\alpha$, revealing a
non-isotropic gas distribution.  
In line with our previous results on \mgii\ kinematics suggesting the presence of outflows in G1, {at minimum}  
a simple 3-D static double-cone model (to represent the trace of bipolar outflows) is required to recreate the EW spatial distribution. 
{The $D$ and $\alpha$ values probed by the arc cannot confirm the presence of a disc, but the data highly disfavor a disc alone.} Our results 
support the interpretation that the EW-$\alpha$ correlation observed
statistically using other extant probes is partly shaped by bipolar metal-rich winds. 
\end{abstract}

\begin{keywords}
galaxies: evolution --- galaxies: formation --- galaxies: intergalactic medium
--- galaxies: clusters: individual (\sgas)
\end{keywords}

\section{Introduction} 
\label{introduction}

The galactic-scale kinematics and spatial structure of the
high-redshift circum-galactic medium~\citep[CGM; ][and references
  therein]{Tumlinson2017, Peroux2020} is an open topic in our
understanding of the baryon cycle of galaxies throughout cosmic
time.
{ The cool ($T\sim 10^4$ K) CGM  is predicted to have an azimuthal dependence
due to the orientation of the material with respect to the central
galaxy: galactic outflow signatures are expected to be more prominent
along the galaxy's minor axis, while accretion and signatures of
extended co-rotating discs may be more readily observable along the
major axis~\citep{Stewart2013, DeFelippis2020,Mitchell2020,
  Nelson2020, Fielding2022}. Observationally, such predictions have
been addressed statistically using \mgii\ quasar 
~\citep[e.g., ][]{Bouche2012, Kacprzak2012,Lan2018,Martin2019} and galaxy~\citep[e.g., ][]{Bordoloi2011,Rubin2018a} absorption systems.}

{ On the other hand, spatially resolving the CGM of {\it individual} galaxies is harder due to the paucity of bright background sources. \mgii\ in emission has been detected around star-forming galaxies~\citep{Burchett2021, Zabl2021,  Shaban2021, Leclercq2022,Rupke2019}, but 
    only in the inner CGM, owing to
the emission measure being proportional to density squared.}

{ The only opportunity to resolve the extended, diffuse cool CGM is
  through lensed
  quasars~\citep[e.g.,][]{Rauch2001,Lopez2007,Zahedy2016}, via
  projected quasars/galaxies~\citep{Peroux2018,Zabl2020} or, more
  recently, using giant gravitational arcs~\citep[hereafter
    ''arc-tomography''; ]
  []{Lopez2018,Lopez2020,Mortensen2021,Tejos2021,Bordoloi2022}.} {
  In particular, arc-tomography maximizes the number of spatially
  independent probes per halo and provides} an excellent match to CGM
scales of up to $\sim 100$ kpc.

\citet[][hereafter ``Paper
  I'']{Tejos2021} presented arc-tomography data of an isolated
star-forming galaxy at $z=0.77$ towards \sgas, called G1, { and
  focused on its} CGM kinematics.  Paper I showed that
\mgii\ absorption velocities comply with an extended rotating disc
(hereafter ERD; { see their Figure 5}), implying that part of the cool gas is co-rotating
with the inner ionized interstellar gas, similar to what some quasar
absorber studies have suggested
\citep[][]{Charlton1998,Steidel2002,Chen2014,Ho2017,Rahmani2018,Zabl2019}.
Besides rotation, it presented evidence of out-flowing material from
blue-shifted (with respect to systemic) velocity components towards
the arc and on top of G1 itself.

In this work we follow up on Paper I to test orientation effects on
\mgii\ absorption equivalent width (EW) around { G1}. In the first
part, based solely on observed quantities, we show that the EW spatial
distribution is non-isotropic.  In the second part, inspired by Paper
I results, we use a { 3D} toy model for the spatial distribution of
EW and find that { both a disc and a double cone (that mimics a
  galactic wind) are } required to fit the EW data.

\section{Data}

\begin{figure}
  \centering
  \includegraphics[width=\columnwidth]{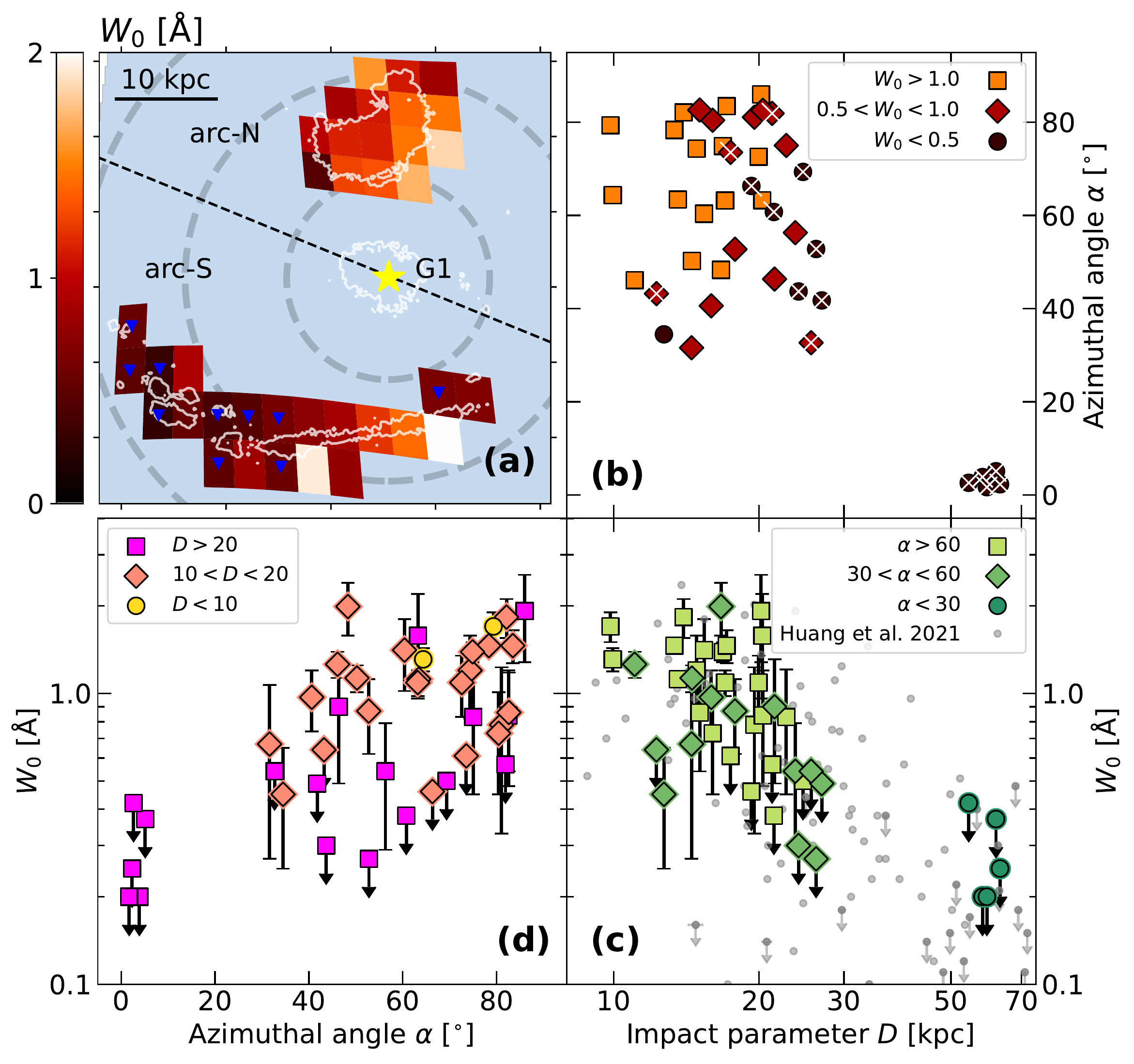}  
  \caption{MUSE data of \sgas\ with G1 at $z=\zabs$. Panel (a): rest-frame
    equivalent-width map around G1 (whose position is marked by a yellow
    star) obtained from $4\times 4$ binned spaxels and shown here in the
    de-lensed absorber plane. Blue arrows indicate 2-$\sigma$
    upper bounds. The dashed straight line indicates G1's major axis (Paper
    I), which we use here to define the azimuthal angle, $\alpha$. The
    concentric circumferences mark impact parameters $D=10$, $20$, and $30$
    kpc. Panels (b)-(c)-(d): the three possible 2-D projections of the
    ($W_0,D,\alpha$)-space, color-coded by the remaining parameter according to
    cuts indicated in the panel labels.  The five positions at $D\approx 50$-$70$ kpc and
    $\alpha\approx 0$ (all non-detections) correspond to arc-E (not displayed in
    Panel (a)). { In panel (b) non-detections are indicated with white crosses.} In panel (c) grey symbols indicate
    quasar absorber measurements~\citep{Huang2021} with impact parameters normalized to
    G1's halo radius.  
    } \label{fig_data}
\end{figure}

The data used in this work were presented in Paper I, and details on
the observations, data reduction, data analysis, and the properties of
G1 are provided therein. A brief summary is given below.

{ The giant arc \sgas~\citep{Koester2010} is produced by (at least)
  one $z = 2.9233$ galaxy, lensed by a massive cluster at $z =
  0.43$. We observed this field }
between April 2018 and January 2019 using the Multi-Unit
Spectroscopic Explorer~\citep[MUSE; ][]{Bacon2010} at the Very Large
Telescope. The observations were carried out in Wide-Field mode with Adaptive
Optics. The data reduction was performed using the MUSE
pipeline~\citep{Weilbacher2012} and 
residual sky contamination
was removed using the Zurich Atmosphere Purge code~\citep{Soto2016}.
The total integration time of the reduced datacube is 3.6~{ hours}, and the effective
PSF FWHM is $0.7^{\prime\prime}$ in the V-band.

The gravitational-arc spectra were optimally averaged leading to binned spaxels
of $0.8^{\prime\prime}\times0.8^{\prime\prime}$ in size in order to (a) increase the signal-noise ratio (S/N) and
(b) minimize seeing-induced cross talk between spaxels.
In the following, these binned spaxels will be referred to as ``positions''. \sgas\ provides 
background light to detect \mgii\ at $z=\zabs$ in 42 such positions, out of which
27 resulted in significant \mgii\ detections and 15 non-detections {(at 2-$\sigma$ confidence)}. { A map of \mgii\ spectra is shown in Figure 3 of~\citet{Tejos2021}}.

\section{Model-independent results}

A map of \mgii\,$\lambda2796$ rest-frame EW ($W_0$) 
is displayed in
Fig.~\ref{fig_data} (a). Each position in the reconstructed (``de-lensed'')
absorber plane defines an impact parameter, $D$, and an azimuthal angle,
$\alpha$. $D$ is defined as the projected distance between spaxel centers and G1's light-weighted barycenter;
$\alpha$ is the angle between G1's major axis and a line connecting G1 and the spaxel centers, with $\alpha=0^{\circ}$ and
$\alpha=90^{\circ}$ corresponding to the major and minor axes, respectively.

{ G1's position and inclination angles (PA~$=68\degree$ and $i=
48\degree$, respectively) are adopted from the ERD model introduced in
Paper I, from which further properties of G1 are: 
  star-formation rate of $1.0 \pm 0.2$\,\msun\,yr$^{-1}$, and halo mass of 
  $10^{11.7 \pm 0.2}$ \msun.}

\sgas\ consists of three bright arcs, named arc-N, arc-S, and arc-E
(only the first two are shown in Fig.~\ref{fig_data} (a); for arc-E, see Paper I).
As seen in panels (b), (c), and (d) { of Fig.~\ref{fig_data}}, these arcs probe
G1's CGM at $\alpha$ uniformly sampled between $\approx30^{\circ}$ and
$\approx90^{\circ}$ at rather similar impact parameters $D\approx
10$--$30$~kpc. Compared with previous arc-tomography data, this
configuration has an edge to test outflow scenarios along the minor
axis, { although unfortunately } G1 inclination angle  
is rather low.

\begin{table}
\centering
\caption{\mgii\ statistical properties towards \sgas}
\begin{tabular}{ccccc}
\hline
\hline
(1) & (2) & (3)&\multicolumn{2}{c}{(4)}\\
$D$ & $\langle W_0\rangle,\sigma_W$&
$r_{\alpha W},p$ &\multicolumn{2}{c}{$\langle \kappa\rangle$}
\\
kpc & \AA & &$\alpha>45^{\circ}$& $\alpha<45^{\circ}$\\
\hline
{\Large \strut} 0--20  & $(1.13,0.38)^a$ &$(0.33,0.15)^a$  &$0.90_{-0.09}^{+0.05}$ & $0.75_{-0.25}^{+0.15}$\\
20--30 & $(0.87,0.48)^a$ &$(0.43,0.40)^a$  &$0.60_{-0.16}^{+0.14}$ & $0.00_{-0.00}^{+0.25}$\\
\hline
\label{table_kappa}
\end{tabular}
\vspace{-0.5cm}
\flushleft (1) Impact parameter; (2) EW median and standard deviation; (3)
EW vs. $\alpha$ Pearson correlation {corefficient} and corresponding two-tailed $p$-value;(4) Covering fraction for $W_0>0.3$ \AA\ and 1-$\sigma$ errors.\\
$^a$ Detections only.
\end{table}

\subsection{Impact parameter dependence}

Fig.~\ref{fig_data} (c) shows $W_0$ versus $D$. The
usual anti-correlation seen in quasar
absorbers~\citep[e.g.,][]{Chen2010,Nielsen2013,Huang2021} is observed, but
thanks to our arc-tomography technique we see it here around an {\it individual} galaxy { at intermediate redshift.} 
Having this spatial information
reduces biases introduced from heterogeneous halo masses and mis-assignments in
galaxy-QSO pair samples~\citep[e.g., ][and references therein]{Ho2020}.

For comparison, quasar absorbers~\citep[][grey symbols]{Huang2021} are displayed, with impact parameters normalized to
G1 halo radius ($123$ kpc; Paper~I).  The { EW} scatter in the quasar sample ($0.46$ and $0.49$
\AA\ at $0<D<20$\,kpc and $20<D<30$\,kpc, respectively) is comparable with the scatter
around G1 (Table~\ref{table_kappa}). This differs from  previous
arc-tomography results~\citep{Lopez2018,Lopez2020}, where the { EW} scatter is
significantly lower than in quasar absorbers. An assessment of this difference, though, is beyond the scope of 
this paper.


\subsection{Azimuthal angle dependence}
\label{sect_azimuthal}

Panel (d) of Fig.~\ref{fig_data} 
shows $W_0$ versus $\alpha$. { Splitting into ranges of $D$,}  a mild trend is { detected} of
stronger systems aligned with the minor axis in both $D$ ranges using a Pearson
correlation test ({ Null-hypothesis rejected at just $\approx 1$-$\sigma$ confidence; }Table~\ref{table_kappa}).  However, this correlation
may be affected, if not driven, by the $W_0-D$ anti-correlation
combined with the particular arc orientation with respect to the
galaxy.

While not included in the statistical test, this potential
$W_0$-$\alpha$ correlation is supported by the lack of detections
along the major axis (discussed in~\S~\ref{sect_anisotropy}) combined
with strong absorption along the minor axis.  Indeed, there is a clear
paucity of weak systems ($W_0<0.5$ \AA) along the minor axis
($\alpha\ga 70^{\circ}$), suggesting some geometrical effect. We note
most of these measurements come from spaxels along the West-side of
arc-N, where the S/N is highest.  We conclude that there is likely an
azimuthal dependence on the measured { EW} across the arc, {although our statistical tests are indecisive}.

An azimuthal effect around G1 {would be} consistent with { \mgii\ averages around star-forming galaxies at similar impact parameters ($D<40$\,kpc), using stacked spectra} of either quasars~\citep{Lan2018} or
galaxies~\citep{Bordoloi2011}. Unfortunately, due to the lack of spaxels with
$\alpha<30^{\circ}$ in this range of $D$, we cannot test claims that \mgii\ { EW}
is bimodal in $\alpha$~\citep{Bouche2012,Kacprzak2012,Martin2019}.

\subsection{Anisotropy and covering fraction}
\label{sect_anisotropy}

Fig.~\ref{fig_data} (a) suggests that spaxels towards arc-S have not
only lower $W_0$ values in general but also a higher fraction of
non-detections than spaxels towards arc-N. In the $10<D<20$\,kpc ring,
for instance, this cannot be due to different S/N levels, given the stringent 
(2-$\sigma$) upper limits on arc-S.  Furthermore, arc-S detections are 
concentrated along the minor axis. Overall, these trends imply an 
anisotropic distribution of the enriched cold gas around G1 at distances
$\approx 1/5$ of the virial radius.

Thanks to arc-tomography, we can assess this effect quantitatively through the
\mgii\ gas covering fraction, $\langle\kappa\rangle$, obtained from a binomial
distribution of detections and non-detections~\citep{Chen2010}.
Table~\ref{table_kappa} displays $\langle\kappa\rangle$ using a $W_0>0.3$ \AA\ cutoff in two bins of $D$ and $\alpha$.  The bins in $D$ exclude the 5 large $D$ (and
low $\alpha$) arc-E non-detections. The bins in $\alpha$ are arbitrarily split
at $45^{\circ}$, which, although resulting in samples of different sizes (6
``major-axis'' spaxels and 29 ``minor-axis'' spaxels), removes the 
selection function { introduced by the particular arc/absorber geometry}. $\langle\kappa\rangle$ appears larger towards the minor axis in both low and high impact parameter bins ($\approx$ 1-$\sigma$ and 3-$\sigma$ significance, respectively).

Combining all the spaxels (i.e., $D<30$\,kpc),
$\langle\kappa\rangle=0.80_{-0.08}^{+0.06}$ (for $\alpha>45^\circ$) and
$0.43_{-0.17}^{+0.18}$ ($\alpha<45^\circ$).  These figures suggest that directions closer to the minor
axis have higher covering fraction than those at $\approx 30^\circ$ from the major-axis, supporting
more clumpiness along the latter, i.e., { suppresion of \mgii} on kpc scales.  

This minor axis $\langle\kappa\rangle$ is coincident with quasar absorber values~\citep{Kacprzak2012,Huang2021} 
(for $D < 40$\,kpc and same $W_0$ cutoff) 
around isolated star-forming galaxies.  On
the other hand, a $\langle\kappa\rangle$-$\alpha$ {\it correlation} like the
one found here is not significantly seen in the~\citet{Huang2021} sample, or
is simply different than that in the~\citet{Kacprzak2012} sample (where
$\langle\kappa\rangle$ peaks both at high {\it and} low $\alpha$).  While we
could elaborate on how sample selection affects these apparent mismatches, we
caution that due to individual spaxels having a spatial extent our measurements are
intrinsically different from $\langle\kappa\rangle$ measured towards point-source quasars;
thus, both measurements are not directly comparable, 
at least using the same $W_0$ cutoff. 

\section{Toy model of the gas distribution}
\label{sect_model}

To help interpret the present $(W_0,D,\alpha)$ data, and inspired by Paper I ERD model, we attempt a 3-D toy
model for the spatial distribution {\it of EW only}, i.e., we do not include
the effects of kinematics or clumpiness.

\subsection{Model parameters and MCMC simulations}

\begin{figure}
  \includegraphics[clip, 
  width=1\columnwidth]{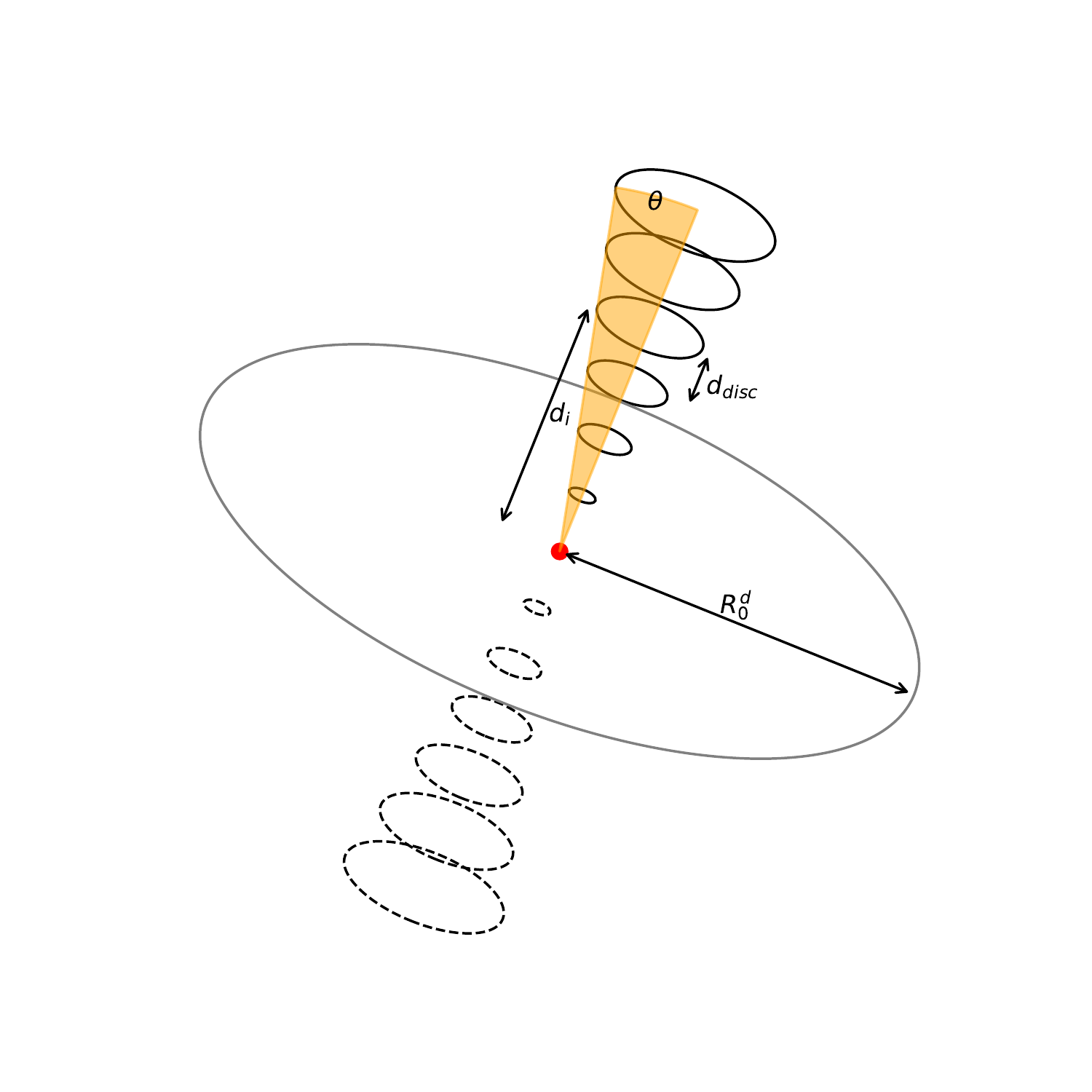}  \\
  \caption{ Model schematic}
\label{fig_schematic}
\end{figure}

Our model consists of a main (static) disc and a double cone that mimics the
possible trace of a biconical
wind~\citep{Shopbell1998,Heckman2000,Ohyama2002,Bouche2012,Schroetter2019}. Both produce a smooth { EW}
distribution on the plane of the sky.

{ A schematic of the model is found in Fig.~\ref{fig_schematic}. }The main disc is an inclined, infinitely
thin disc at the position of G1. PA and inclination are adopted from Paper~I
ERD model. On the main disc EW is a function of distance  to its center ($R$) only: $W_0 =
W_0^d\exp{(-R/R_0^d)}$, where $W_0^d$ and $R_0^d$ are maximum EW and
characteristic radius, respectively.

The double cone is implemented by stacking $n_{\rm disc}$ parallel and
concentric discs on either side of the main-disc along its axis of symmetry. 
These discs have radii
$d_i\tan{\theta}$, where $\theta$ is the half opening angle of a cone and
$d_i$ the distance of the $i$-th disc to the center of G1 along the axis of
symmetry. { The discs} are separated by a constant distance $d_{\rm disc}$ from each
other. For ease of implementation, each ``truncated'' disc contributes with a
constant EW, i.e., $W_0 = W_0^c$.  

The total synthetic EW is the sum of both
contributions along the line-of-sight and is evaluated at a given RA-DEC
coordinate.

\begin{table}
\caption{MCMC parameter priors}
\label{tab:Priors}
\begin{tabular}{cccc}
\hline
\hline
Parameter & Min. Value & Max. Value\\
\hline
$W_0^d$ [\AA]& 0.0 & 5.0\\
\vspace{4pt}
$R_0^d$ [kpc] & 0.0 & 100.0\\
\vspace{4pt}
$\theta$ [$^\circ$] & 0.0 & 90.0\\
\vspace{4pt}
$W_0^c$ [\AA] & 0.0 & 0.2\\
\hline
\end{tabular}
\end{table}

We perform Monte Carlo Markov Chain (MCMC) simulations on the (RA-DEC,EW) data in order
to (a) find representative parameters for comparing models and (b) study
degeneracies in the model given our data.  The following four model parameters
are considered $W_0^d$, $R_0^d$, $\theta$ and $W_0^c$. {For these four parameters we assume uniform priors  between the minimum and maximum values listed in Table \ref{tab:Priors}. For the disc-only model the cone model parameters are fixed to 0, and vice versa for the cone-only model.}  The rest {of the model parameters} are fixed at
$n_{\rm disc}=25$ and $d_{\rm disc}=1$ kpc, i.e., cones extend out to $25$ kpc
North and South of G1 ($n_{\rm disc}$ and $d_{\rm disc}$ values are less sensible provided their product is constant). { The assumed likelihood function (${\rm \mathcal{L}}$) for spaxels with detected absorption is given by:

\begin{equation}
\label{eq:Like}
\begin{aligned}
\rm log\mathcal{L} &  = \sum_{i} \frac{-(W_{0,i} - W_{model,i}(D_{i},\alpha_{i}))^2}{2\sigma_{i}^{2}}  -0.5log(2\pi \sigma_{i}^{2})\\
& + n_{con}log(0.954) + n_{incon}log(0.046), 
\end{aligned}
\end{equation}

\noindent where ${\rm W_{0,i}}$ and ${\rm \sigma_{i}}$ are the detected rest-frame equivalent widths and errors for a given spaxel, and ${\rm W_{model,i}(D_{i},\alpha_{i}) }$ is the proposed modelled rest-frame equivalent width at the position of the observed spaxel. To account for spaxels with $2\sigma$ EW upper limits in the likelihood, we include the probability each spaxel is consistent (95.4 per cent) or inconsistent (4.6 per cent) with the proposed model in the likelihood function to represent the $2\sigma$ confidence in the limits. Thus in Equation \ref{eq:Like}, $n_{con}$ and $n_{incon}$ represent the total number spaxels with EW upper limits that are consistent and inconsistent (respectively) with the proposed model.}

\subsection{Results and discussion}

\begin{figure*}
  \centering
  \subfloat{\includegraphics[width=0.9\columnwidth]{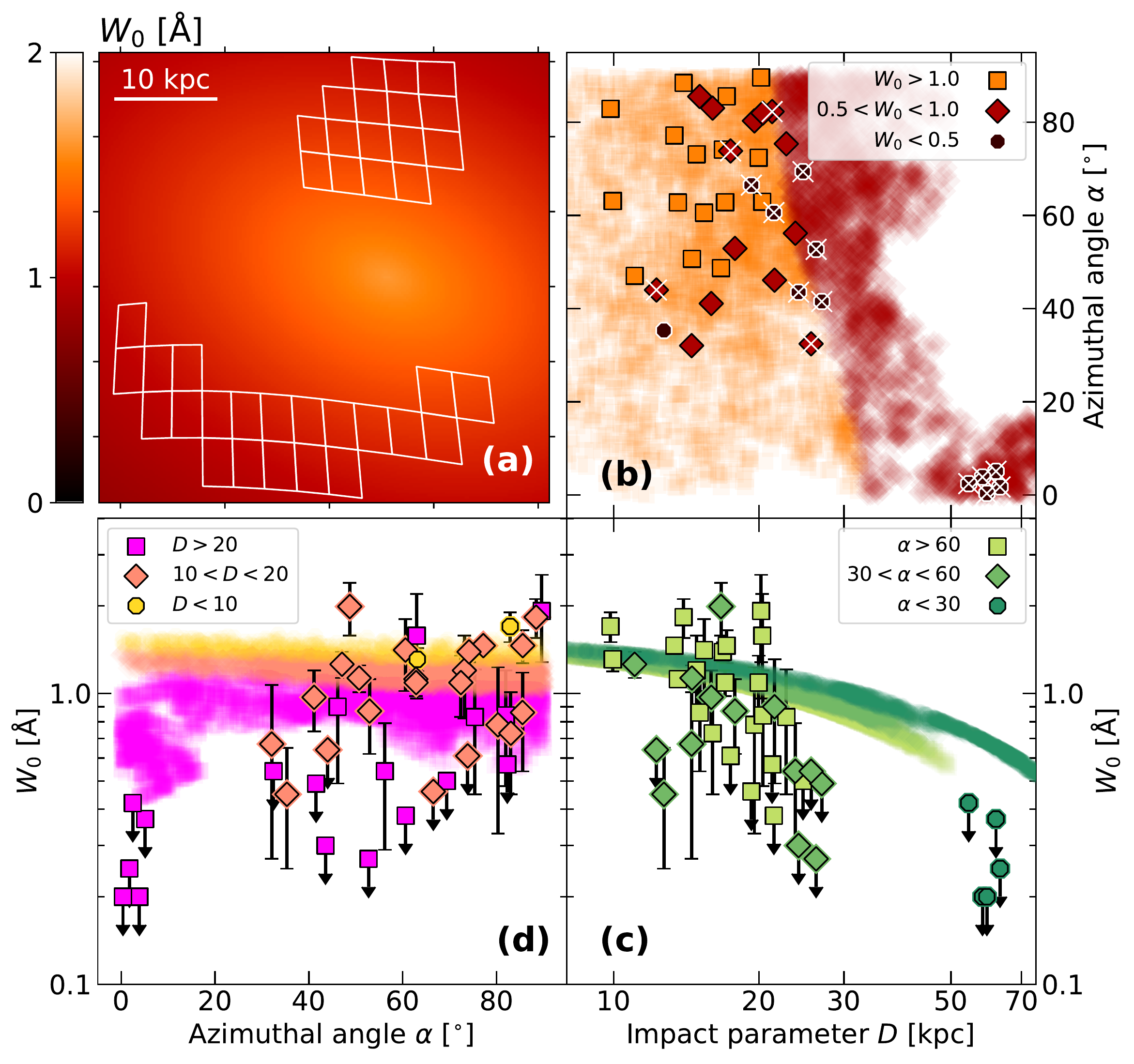} } 
  \subfloat{\includegraphics[width=0.7\columnwidth]{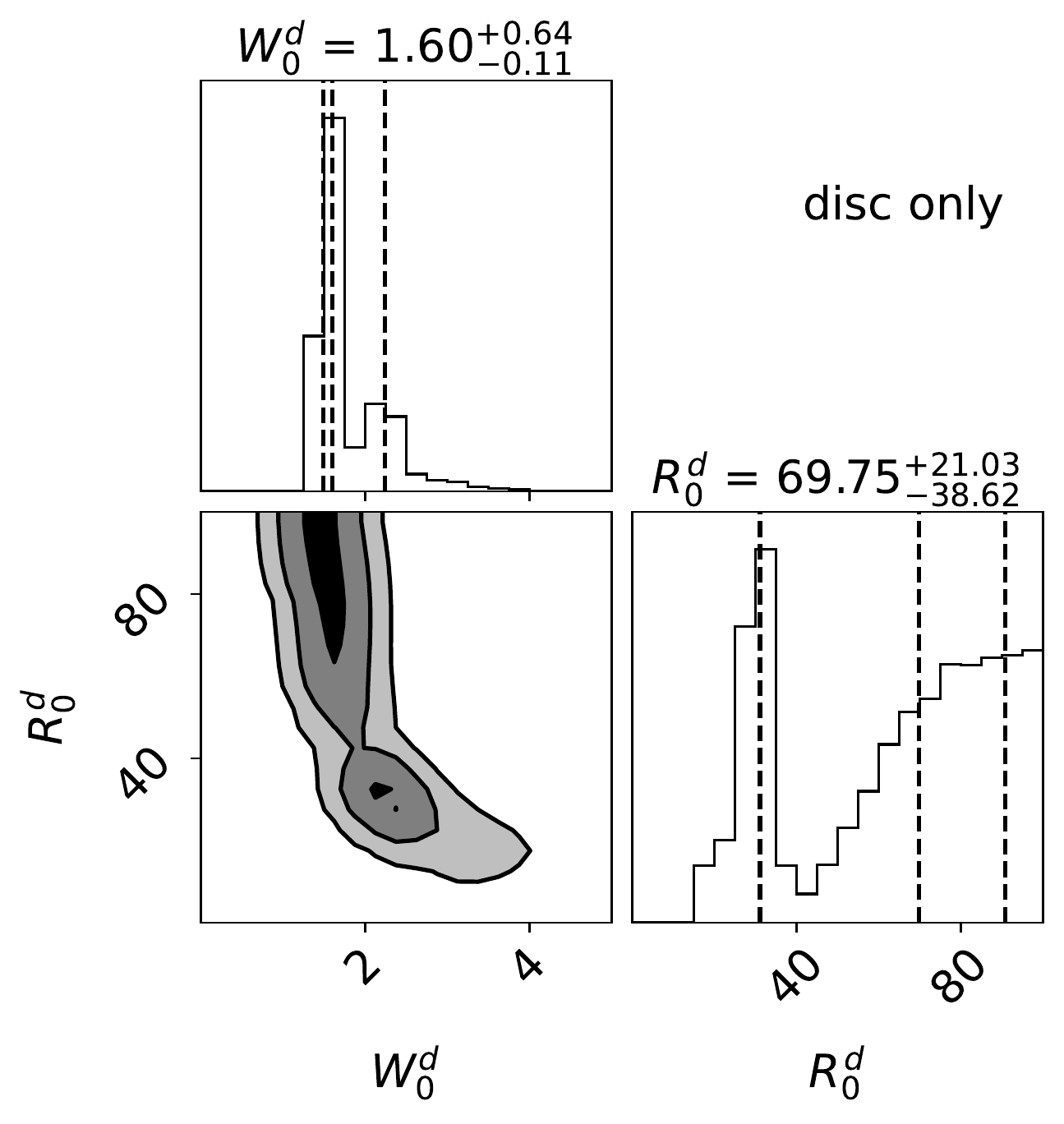} }\\
    \subfloat{\includegraphics[width=0.9\columnwidth]{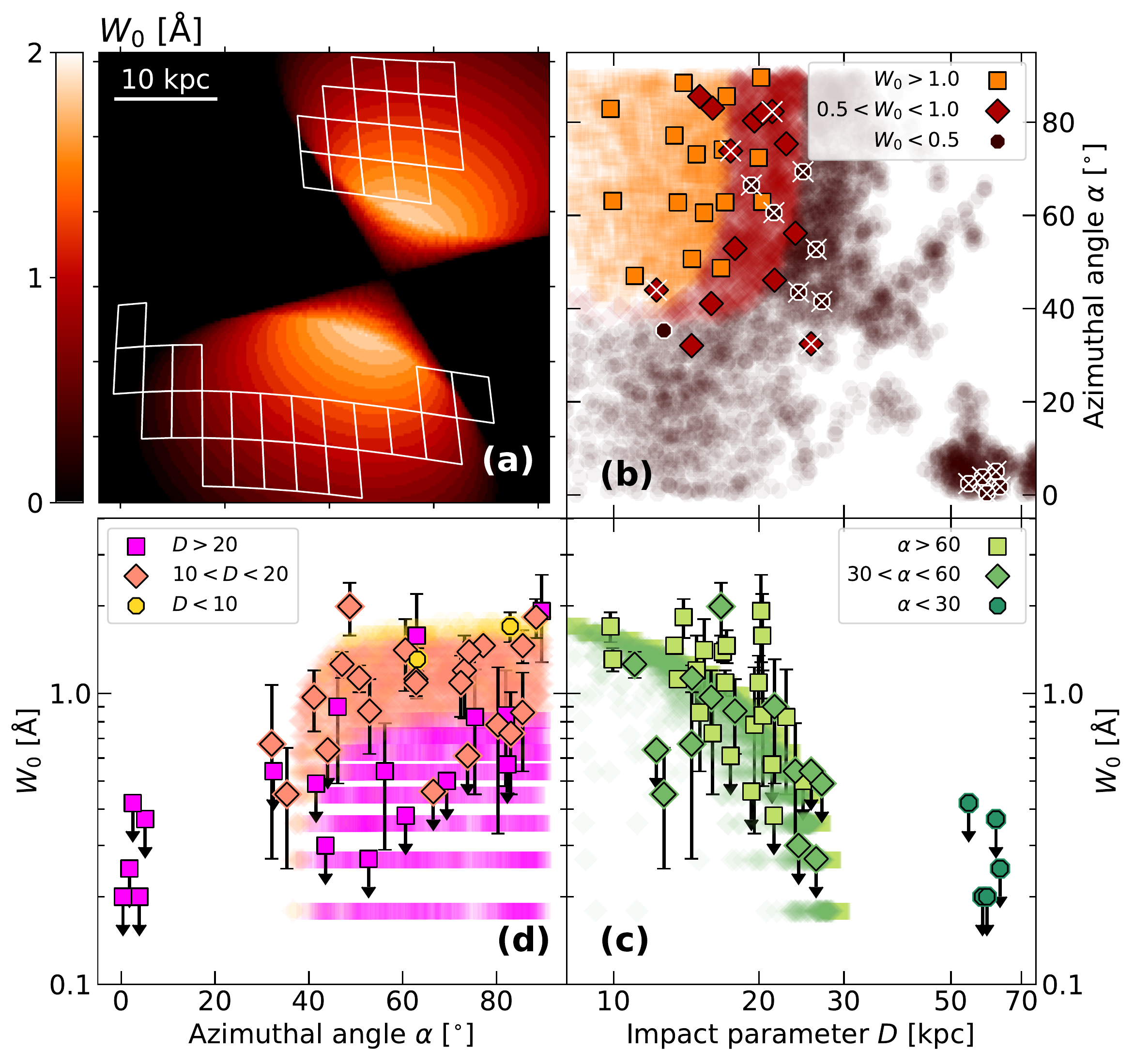} } 
  \subfloat{\includegraphics[width=0.7\columnwidth]{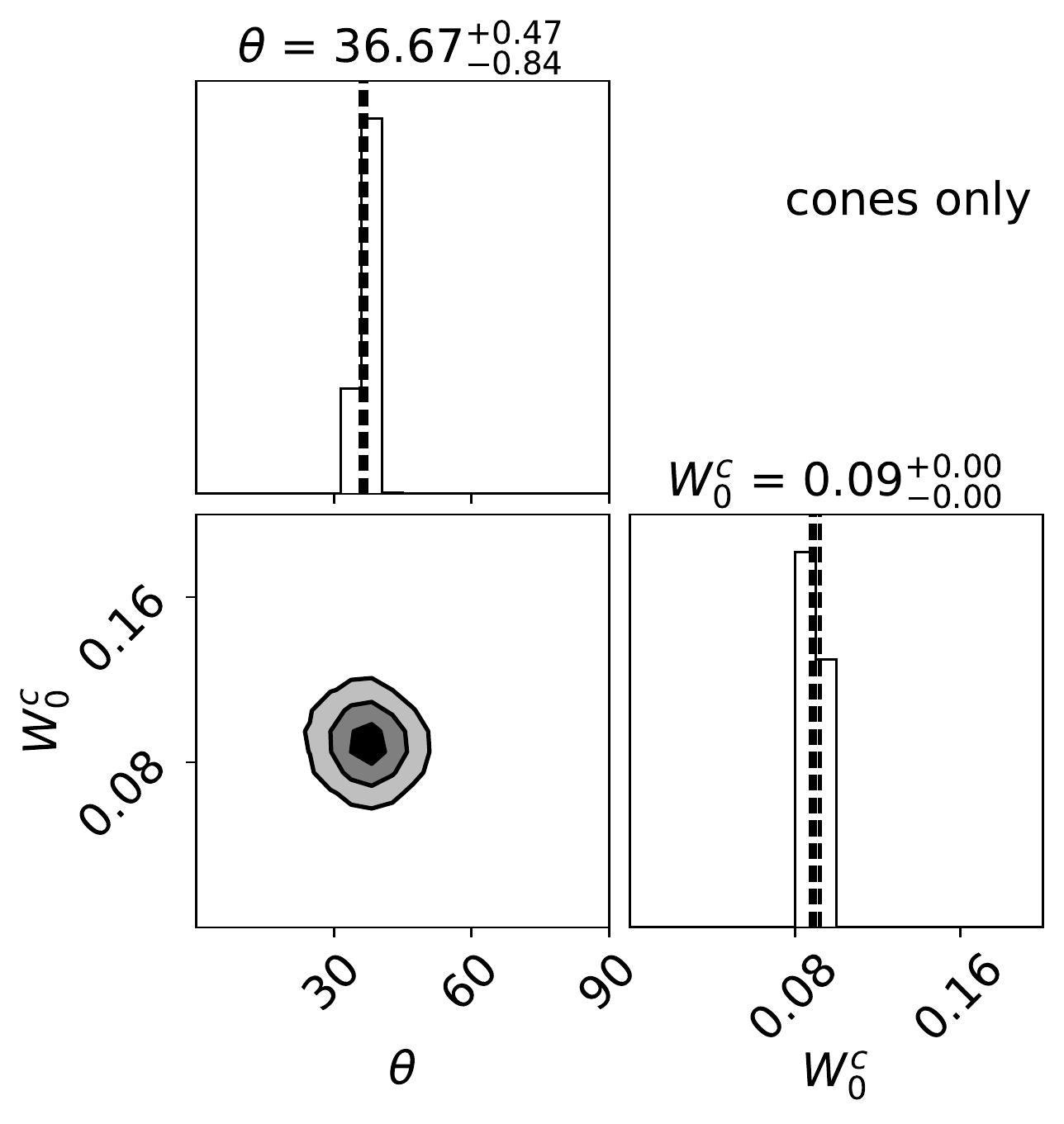} }\\
  \subfloat{\includegraphics[width=0.9\columnwidth]{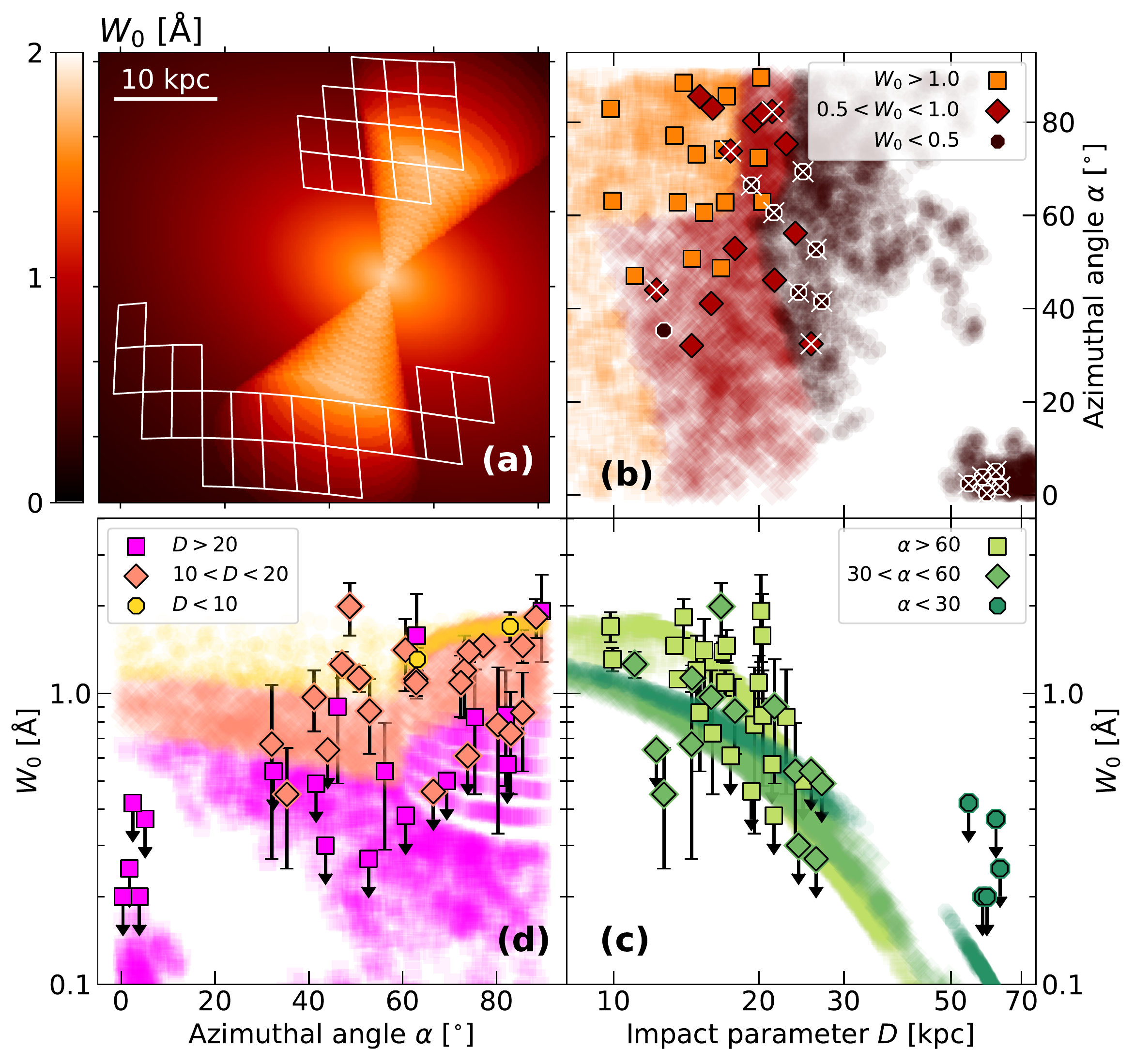} } 
  \subfloat{\includegraphics[width=0.8\columnwidth]{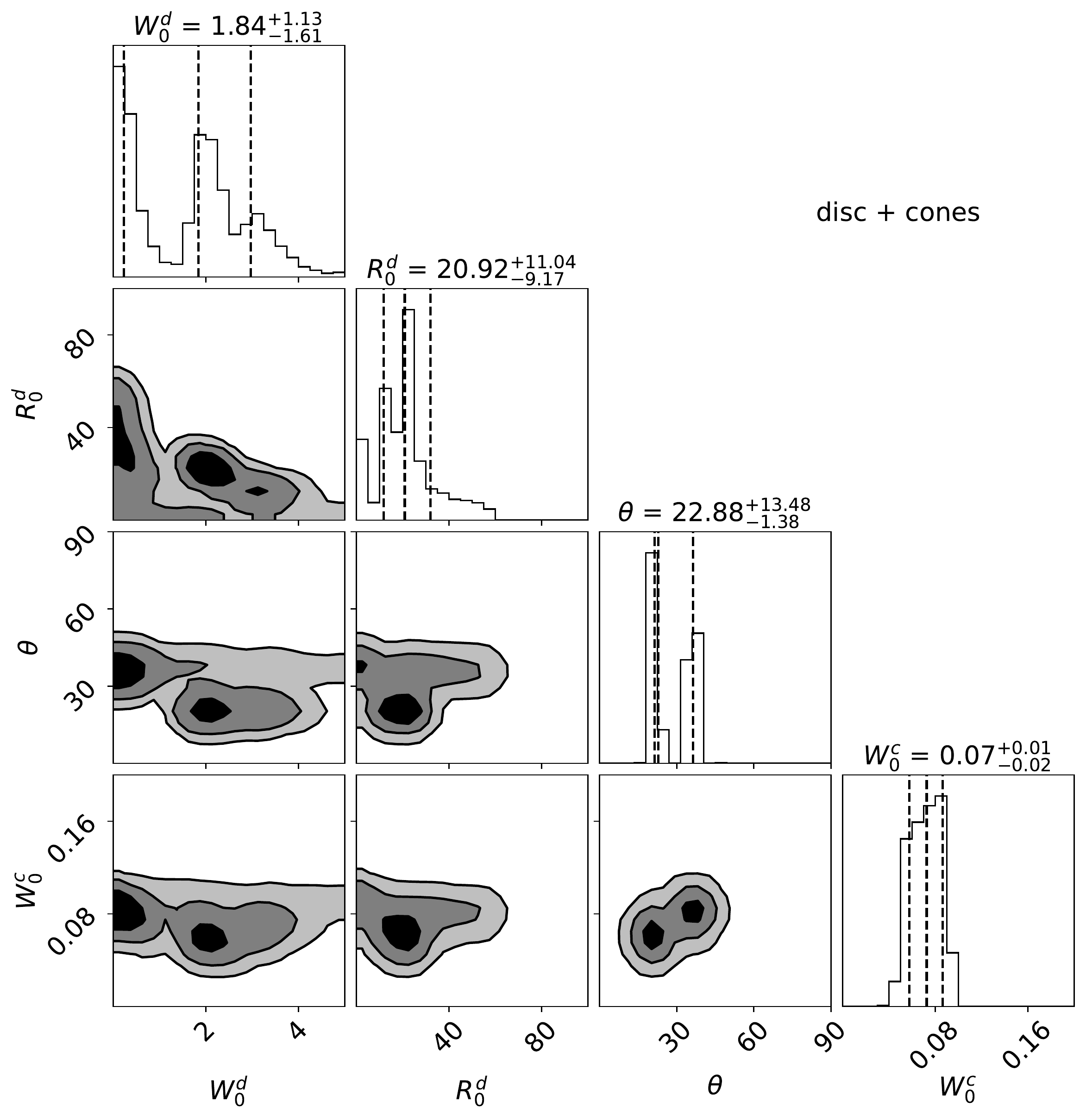} }
  \caption{ {\it Left-hand column:} same as Fig.~\ref{fig_data} but adding the EW models `disc-only',
    `cones-only', and `disc+cones' (top to bottom).  (a) panels display the synthetic EW maps, with $W_0$ 
    evaluated in the de-lensed plane at $0.03^{\prime\prime}$
    sampling. White rhomboids correspond to the spaxel grid shown in
    Fig.~\ref{fig_data}.  Panels (b)-(c)-(d) display the three possible 2-D projections of the ($W_0,D,\alpha$)-space { for data and model}. Data points  have 1-$\sigma$ error bars and black edges, exactly as in Fig~\ref{fig_data}. { In panel (b) non-detections are indicated with white crosses.} The model assumes the median parameter values from the MCMC simulations. $W_0$, $D$ and $\alpha$ are
    evaluated at 200 random positions within $\pm 1$ kpc of each spaxel
    center {(colored fuzzy points)}.   {\it Right-hand column:} corresponding corner plots from the MCMC simulations. }
\label{fig_model}
\end{figure*}

\begin{table}
\centering
\caption{MCMC results}
\begin{tabular}{lccccc}
\hline
\hline
(1)&(2)&(3)&(4)&(5)&(6)\\
  & $W_0^d$ & $R_0^d$ & $\theta$ & $W_0^c$ &BIC\\
            & [$\AA$] & [kpc] & [$^\circ$] & [\AA] & \\
\hline
D  & 
$1.60_{-0.11}^{+0.64}$
& 
$69.75_{-38.62}^{+21.03}$
&
0$^b$ & 0$^b$ & $77^{+1}_{-1}$\\
C  & 
0$^a$ & 0$^a$  
& 
$36.67_{-0.84}^{+0.47}$
& 
$0.09_{-0.00}^{+0.00}$
&$34^{+1}_{-1}$\\
D+C  & 
$1.84_{-1.61}^{+1.13}$
&
$20.92_{-9.17}^{+11.04}$
&
$22.88_{-1.38}^{+13.48}$
&
$0.07_{-0.02}^{+0.01}$
&$40^{+4}_{-2}$\\
\hline
\label{table_mcmc}
\end{tabular}
\vspace{-0.5cm}
\flushleft (1) Model (D: disc-only; C: cones-only; D+C: disc+cones);\\
(2) maximum EW on disc; (3) disc's characteristic radius;\\
(4) Half opening angle; (5) constant EW on cone discs; \\
(6) Median Bayesian information criterion with $\pm25$-percentile errors.\\ 
$^{a,b}${Parameter fixed to 0 to exclude disc or cone component}.
\end{table}

{ Setting the appropriate parameters to zero, the model enables 3
  flavors: ``disc-only'', ``cones-only'', and ``disc+cones''.}
{For each flavour of model, Table~\ref{table_mcmc} displays the
  median parameter values from the MCMC with 68\% confidence limits
  along with the median and interquarile range of the Bayesian
  information criterion.}  { We use these results to compare
  the different model flavors qualitatively and generate the synthetic
  EW maps and projections shown in Figure~\ref{fig_model} (left-hand
  column). The corner plots for each of the three model types are
  provided in the right-hand column of the figure.}
    
    { For the disc-only model, there appears to be two preferred
      parameter sets in the respective corner plots (top right of
      Figure \ref{fig_model}), and an apparent degeneracy between
      $W_0^d$ and $R_0^d$. 
      More constraints along the
      major axis would be needed to break the degeneracy. The
      cones-only model parameter space is well constrained by the data
      (middle right panel of Figure \ref{fig_model}). When comparing
      to the single component models to the disc+cone model (bottom
      right of Figure \ref{fig_model}), we note that the median value
      of the disc $R_0^d$ decreases significantly to reduce the EW
      contribution from the disc component. However, there are two
      preferred parameter sets. One set is identical to the cone-only
      model with no disc contribution, while the other set prefers a
      small disc in addition to cones with a smaller opening angle
      ($\theta\approx23^\circ$). Based on the Bayseian information
      criterion (Table \ref{table_mcmc}), the data marginally prefers
      the cones-only model over a disc+cones model. The disc-only
      model is a poor description of the data.}
  
  We warn that, given the lack of data along the major axis, the data are not constraining enough to remove degeneracies in the model parameter space. In particular, the data cannot accurately account for the contribution of the disc at $D\la 30$ kpc in models that include a disc component. Thus we can only use these as toy models.  With this disclaimer in mind,  { a qualitative comparison between data and  each model flavor is as follows:}

\begin{enumerate}
\item {\it Disc-only model}: The $\alpha$-$D$ projection (Panel b)
   is not as well
  reproduced visually as for models including cones.  {$W_0$-$D$ (c)  also offers a poor fit to the data, showing almost no 
  $\alpha$-driven scatter}. 
  $W_0$-$\alpha$ (d) is definitively not well reproduced, at least for $D>10$ kpc.  
\item {\it Cones-only model}: $\alpha$-$D$ seems better reproduced, with a
  tapered vertical gradient in $W_0$ (panel b).  A scatter in EW emerges in the $W_0$-$D$ projection (c) as a
  consequence of more anisotropy and an $W_0$-$\alpha$ (d) correlation is
  recovered, which matches the data reasonably well.

\item {\it Disc+cones model}: All three projections are at least as well matched as for the cones-only model.  Low-$(D,\alpha)$ measurements are not tied to the disc part of the model, although this might be due to the lack of minor axis measurements at $D<30$ kpc.

\end{enumerate}

We conclude that, {at minimum, a ``cone-dominated'' component is necessary to provide a better 
description of the present data than a disc-only model}. { Requiring a cone component suggests} that the
observed EW scatter in $(W_0$-$D)$ is driven by anisotropy.

As demonstrated in Paper I, the kinematic data of our system supports an ERD model. 
However, the EW data presented in this work suggests a more
complex model. This apparent discrepancy is likely a result of how the
kinematic information is derived, as the MUSE velocity centroids are likely
dominated by the highest column density clouds.  If these were preferentially
located on the disc mid-plane, which Paper I and our model idealize as a thin
disc, both kinematics and EW would match. Conversely, if the dominant clouds
are distributed symmetrically off the mid-plane, and still entrained by
rotation (i.e., a thick disc), then kinematics would be well fitted but EW
not, because the line of sight would miss some of the off-disc clouds. This
latter situation is possible if galactic-scale outflows are
present~\citep{Martin2012,Rubin2014} or velocity dispersion is high, 
{ the latter of} which has been suggested for this galaxy~\citep[Paper~I; ][]{Mortensen2021}.

\subsection{Caveats}\label{sect_caveats}

Evidently, a proper model of the CGM must also consider
kinematics~\citep{Martin2019,Afruni2021} as EW is basically a measure of
line-of-sight velocity dispersion.  But perhaps even more fundamental, our
model neglects the physics of winds. Assuming { a biconical outflow made of
constant EW discs} violates mass conservation, which predicts { that  gas density $\propto d_i^{-1}$}~\citep[e.g., ][]{Schroetter2019}.  

{ Regarding observational caveats, }
background light is
assumed to be spatially homogeneous within spaxels, which is most likely
incorrect on kpc scales. Instead, we assume the spaxel central value is a
representative (light-weighted) average, akin to using background
galaxies~\citep[e.g.,][]{Bordoloi2011,Diamond-Stanic2016,Zabl2020}.  To test
robustness we re-ran the simulations with randomized spaxel positions and
verified that results do not change within $\pm 1$ kpc (95\% confidence limit)
of spaxel centers.
Atmospheric effects are neglected as well (although our
aggressive spatial binning should counteract them). { Finally, $D$ and $\alpha$ values are based on the lens model presented in Paper I.}

 Summarizing, our toy
model highlights the power of having spatially-resolved sampling from arc-tomography data
to constrain models of the CGM, but also that CGM models require a lot more
complexity.

\section{Summary and conclusions}

We have analyzed arc-tomography data of \sgas\ at $z=0.77$ to assess
possible orientation effects on \mgii\ rest-frame equivalent width,
$W_0$. The arc configuration is well suited given the uniform sampling
of impact parameters $D$ and azimuthal angles $\alpha$
(Fig.~\ref{fig_data} b) although major axis positions at $D<30$ kpc 
are under-represented.  We have discussed the 3 projections of the
($W_0,D,\alpha$)-space and compared them with QSO absorber statistics
and with a simple disc + double cone model for the $W_0$ spatial
distribution. Our conclusions are spelled as follows:

\begin{enumerate}
\item From observational data alone:
  \begin{enumerate}
  \item $W_0$ and $D$ anti-correlate. The scatter in $W_0$ is comparable with
    quasar-absorber statistics.
  \item $W_0$ and $\alpha$ mildly correlate, which is consistent
    with \citet{Bordoloi2011} and \citet{Lan2018}.
  \item Covering fraction and $\alpha$ correlate,
    suggesting less clumpiness along the minor axis.
  \item {2-D projections of the ($W_0,D,\alpha$)-space are difficult to control by the remaining parameter. } The most {deterministic} diagnostics to assess orientation effects
    is the EW spatial distribution  itself, on which 3-D models of the CGM can
    be tested. 
  \end{enumerate}
\item From a comparison with $W_0$ by model:
  \begin{enumerate}
\item {The data favours a double cone model component, mimicking the trace of a galactic wind. }
\item Both the $\alpha$-$D$ and the $W_0$-$\alpha$ projections are model
  constraining, specially the former, which
  reflects the arc/galaxy configuration. $W_0$-$D$ is the least constraining 
  due to $W_0$ anisotropy, {although it can help to reject some models if $\alpha$ is well sampled}.
\item $\alpha$ seems to be a key parameter in constraining models of the CGM, that so
far can only be done in a less-biased fashion with
arc-tomography.
  \end{enumerate}
\end{enumerate}

Gravitational arc-tomography provides unprecedented opportunities for
assessing the spatial structure of the high-redshift CGM. We look forward to
new arc configurations with appropriate $D$ and $\alpha$ samplings, ideally
intercepting more inclined galaxies, through which we expect orientation effects to
be strongest. {Irrespective of this,  we hope that the current results will help to better interpret single-sightline absorber samples.}
\\

\section*{acknowledgments}
{ We thank the anonymous referee for their careful review and comments that improved the quality of this paper.  We also warmly thank Eric Jullo for discussions on gravitational lensing.}
This paper is based on observations collected at the European Southern
Observatory under ESO programme(s) 0101.A-0364(A) (PI Lopez).  AF, SL, NT, and
MH acknowledge support by FONDECYT grant 1191232.
EJJ acknowledges support from FONDECYT Iniciaci\'on en investigaci\'on
Project 11200263.

\section*{Data availability}
The data may be
accessed from the ESO Archive at  {\tt http://archive.eso.org/}  using the above program ID.

\bibliographystyle{mnras}
\bibliography{Lopez_lit}

\begin{thebibliography}{}
\makeatletter
\relax
\def\mn@urlcharsother{\let\do\@makeother \do\$\do\&\do\#\do\^\do\_\do\%\do\~}
\def\mn@doi{\begingroup\mn@urlcharsother \@ifnextchar [ {\mn@doi@}
  {\mn@doi@[]}}
\def\mn@doi@[#1]#2{\def\@tempa{#1}\ifx\@tempa\@empty \href
  {http://dx.doi.org/#2} {doi:#2}\else \href {http://dx.doi.org/#2} {#1}\fi
  \endgroup}
\def\mn@eprint#1#2{\mn@eprint@#1:#2::\@nil}
\def\mn@eprint@arXiv#1{\href {http://arxiv.org/abs/#1} {{\tt arXiv:#1}}}
\def\mn@eprint@dblp#1{\href {http://dblp.uni-trier.de/rec/bibtex/#1.xml}
  {dblp:#1}}
\def\mn@eprint@#1:#2:#3:#4\@nil{\def\@tempa {#1}\def\@tempb {#2}\def\@tempc
  {#3}\ifx \@tempc \@empty \let \@tempc \@tempb \let \@tempb \@tempa \fi \ifx
  \@tempb \@empty \def\@tempb {arXiv}\fi \@ifundefined
  {mn@eprint@\@tempb}{\@tempb:\@tempc}{\expandafter \expandafter \csname
  mn@eprint@\@tempb\endcsname \expandafter{\@tempc}}}

\bibitem[\protect\citeauthoryear{{Afruni}, {Fraternali}  \&
  {Pezzulli}}{{Afruni} et~al.}{2021}]{Afruni2021}
{Afruni} A.,  {Fraternali} F.,   {Pezzulli} G.,  2021, \mn@doi [\mnras]
  {10.1093/mnras/staa3759}, \href
  {https://ui.adsabs.harvard.edu/abs/2021MNRAS.501.5575A} {501, 5575}

\bibitem[\protect\citeauthoryear{{Bacon} et~al.,}{{Bacon}
  et~al.}{2010}]{Bacon2010}
{Bacon} R.,  et~al., 2010, in Ground-based and Airborne Instrumentation for
  Astronomy III. p. 773508, \mn@doi{10.1117/12.856027}

\bibitem[\protect\citeauthoryear{{Bordoloi} et~al.,}{{Bordoloi}
  et~al.}{2011}]{Bordoloi2011}
{Bordoloi} R.,  et~al., 2011, \mn@doi [\apj] {10.1088/0004-637X/743/1/10},
  \href {http://adsabs.harvard.edu/abs/2011ApJ...743...10B} {743, 10}

\bibitem[\protect\citeauthoryear{{Bordoloi} et~al.,}{{Bordoloi}
  et~al.}{2022}]{Bordoloi2022}
{Bordoloi} R.,  et~al., 2022, \mn@doi [\nat] {10.1038/s41586-022-04616-1},
  \href {https://ui.adsabs.harvard.edu/abs/2022Natur.606...59B} {606, 59}

\bibitem[\protect\citeauthoryear{{Bouch{\'e}}, {Hohensee}, {Vargas},
  {Kacprzak}, {Martin}, {Cooke}  \& {Churchill}}{{Bouch{\'e}}
  et~al.}{2012}]{Bouche2012}
{Bouch{\'e}} N.,  {Hohensee} W.,  {Vargas} R.,  {Kacprzak} G.~G.,  {Martin}
  C.~L.,  {Cooke} J.,   {Churchill} C.~W.,  2012, \mn@doi [\mnras]
  {10.1111/j.1365-2966.2012.21114.x}, \href
  {http://adsabs.harvard.edu/abs/2012MNRAS.426..801B} {426, 801}

\bibitem[\protect\citeauthoryear{{Burchett}, {Rubin}, {Prochaska}, {Coil},
  {Vaught}  \& {Hennawi}}{{Burchett} et~al.}{2021}]{Burchett2021}
{Burchett} J.~N.,  {Rubin} K. H.~R.,  {Prochaska} J.~X.,  {Coil} A.~L.,
  {Vaught} R.~R.,   {Hennawi} J.~F.,  2021, \mn@doi [\apj]
  {10.3847/1538-4357/abd4e0}, \href
  {https://ui.adsabs.harvard.edu/abs/2021ApJ...909..151B} {909, 151}

\bibitem[\protect\citeauthoryear{{Charlton} \& {Churchill}}{{Charlton} \&
  {Churchill}}{1998}]{Charlton1998}
{Charlton} J.~C.,  {Churchill} C.~W.,  1998, \mn@doi [\apj] {10.1086/305632},
  \href {http://adsabs.harvard.edu/abs/1998ApJ...499..181C} {499, 181}

\bibitem[\protect\citeauthoryear{{Chen}, {Helsby}, {Gauthier}, {Shectman},
  {Thompson}  \& {Tinker}}{{Chen} et~al.}{2010}]{Chen2010}
{Chen} H.-W.,  {Helsby} J.~E.,  {Gauthier} J.-R.,  {Shectman} S.~A.,
  {Thompson} I.~B.,   {Tinker} J.~L.,  2010, \mn@doi [\apj]
  {10.1088/0004-637X/714/2/1521}, \href
  {http://adsabs.harvard.edu/abs/2010ApJ...714.1521C} {714, 1521}

\bibitem[\protect\citeauthoryear{{Chen}, {Gauthier}, {Sharon}, {Johnson},
  {Nair}  \& {Liang}}{{Chen} et~al.}{2014}]{Chen2014}
{Chen} H.-W.,  {Gauthier} J.-R.,  {Sharon} K.,  {Johnson} S.~D.,  {Nair} P.,
  {Liang} C.~J.,  2014, \mn@doi [\mnras] {10.1093/mnras/stt2288}, \href
  {http://adsabs.harvard.edu/abs/2014MNRAS.438.1435C} {438, 1435}

\bibitem[\protect\citeauthoryear{{DeFelippis}, {Genel}, {Bryan}, {Nelson},
  {Pillepich}  \& {Hernquist}}{{DeFelippis} et~al.}{2020}]{DeFelippis2020}
{DeFelippis} D.,  {Genel} S.,  {Bryan} G.~L.,  {Nelson} D.,  {Pillepich} A.,
  {Hernquist} L.,  2020, \mn@doi [\apj] {10.3847/1538-4357/ab8a4a}, \href
  {https://ui.adsabs.harvard.edu/abs/2020ApJ...895...17D} {895, 17}

\bibitem[\protect\citeauthoryear{{Diamond-Stanic}, {Coil}, {Moustakas},
  {Tremonti}, {Sell}, {Mendez}, {Hickox}  \& {Rudnick}}{{Diamond-Stanic}
  et~al.}{2016}]{Diamond-Stanic2016}
{Diamond-Stanic} A.~M.,  {Coil} A.~L.,  {Moustakas} J.,  {Tremonti} C.~A.,
  {Sell} P.~H.,  {Mendez} A.~J.,  {Hickox} R.~C.,   {Rudnick} G.~H.,  2016,
  \mn@doi [\apj] {10.3847/0004-637X/824/1/24}, \href
  {http://adsabs.harvard.edu/abs/2016ApJ...824...24D} {824, 24}

\bibitem[\protect\citeauthoryear{{Fielding} \& {Bryan}}{{Fielding} \&
  {Bryan}}{2022}]{Fielding2022}
{Fielding} D.~B.,  {Bryan} G.~L.,  2022, \mn@doi [\apj]
  {10.3847/1538-4357/ac2f41}, \href
  {https://ui.adsabs.harvard.edu/abs/2022ApJ...924...82F} {924, 82}

\bibitem[\protect\citeauthoryear{{Heckman}, {Lehnert}, {Strickland}  \&
  {Armus}}{{Heckman} et~al.}{2000}]{Heckman2000}
{Heckman} T.~M.,  {Lehnert} M.~D.,  {Strickland} D.~K.,   {Armus} L.,  2000,
  \mn@doi [\apjs] {10.1086/313421}, \href
  {https://ui.adsabs.harvard.edu/abs/2000ApJS..129..493H} {129, 493}

\bibitem[\protect\citeauthoryear{{Ho}, {Martin}, {Kacprzak}  \&
  {Churchill}}{{Ho} et~al.}{2017}]{Ho2017}
{Ho} S.~H.,  {Martin} C.~L.,  {Kacprzak} G.~G.,   {Churchill} C.~W.,  2017,
  \mn@doi [\apj] {10.3847/1538-4357/835/2/267}, \href
  {http://adsabs.harvard.edu/abs/2017ApJ...835..267H} {835, 267}

\bibitem[\protect\citeauthoryear{{Ho}, {Martin}  \& {Schaye}}{{Ho}
  et~al.}{2020}]{Ho2020}
{Ho} S.~H.,  {Martin} C.~L.,   {Schaye} J.,  2020, \mn@doi [\apj]
  {10.3847/1538-4357/abbe88}, \href
  {https://ui.adsabs.harvard.edu/abs/2020ApJ...904...76H} {904, 76}

\bibitem[\protect\citeauthoryear{{Huang}, {Chen}, {Shectman}, {Johnson},
  {Zahedy}, {Helsby}, {Gauthier}  \& {Thompson}}{{Huang}
  et~al.}{2021}]{Huang2021}
{Huang} Y.-H.,  {Chen} H.-W.,  {Shectman} S.~A.,  {Johnson} S.~D.,  {Zahedy}
  F.~S.,  {Helsby} J.~E.,  {Gauthier} J.-R.,   {Thompson} I.~B.,  2021, \mn@doi
  [\mnras] {10.1093/mnras/stab360}, \href
  {https://ui.adsabs.harvard.edu/abs/2021MNRAS.502.4743H} {502, 4743}

\bibitem[\protect\citeauthoryear{{Kacprzak}, {Churchill}  \&
  {Nielsen}}{{Kacprzak} et~al.}{2012}]{Kacprzak2012}
{Kacprzak} G.~G.,  {Churchill} C.~W.,   {Nielsen} N.~M.,  2012, \mn@doi [\apjl]
  {10.1088/2041-8205/760/1/L7}, \href
  {http://adsabs.harvard.edu/abs/2012ApJ...760L...7K} {760, L7}

\bibitem[\protect\citeauthoryear{{Koester}, {Gladders}, {Hennawi}, {Sharon},
  {Wuyts}, {Rigby}, {Bayliss}  \& {Dahle}}{{Koester}
  et~al.}{2010}]{Koester2010}
{Koester} B.~P.,  {Gladders} M.~D.,  {Hennawi} J.~F.,  {Sharon} K.,  {Wuyts}
  E.,  {Rigby} J.~R.,  {Bayliss} M.~B.,   {Dahle} H.,  2010, \mn@doi [\apjl]
  {10.1088/2041-8205/723/1/L73}, \href
  {https://ui.adsabs.harvard.edu/abs/2010ApJ...723L..73K} {723, L73}

\bibitem[\protect\citeauthoryear{{Lan} \& {Mo}}{{Lan} \& {Mo}}{2018}]{Lan2018}
{Lan} T.-W.,  {Mo} H.,  2018, \mn@doi [\apj] {10.3847/1538-4357/aadc08}, \href
  {https://ui.adsabs.harvard.edu/abs/2018ApJ...866...36L} {866, 36}

\bibitem[\protect\citeauthoryear{{Leclercq} et~al.,}{{Leclercq}
  et~al.}{2022}]{Leclercq2022}
{Leclercq} F.,  et~al., 2022, \mn@doi [\aap] {10.1051/0004-6361/202142179},
  \href {https://ui.adsabs.harvard.edu/abs/2022A&A...663A..11L} {663, A11}

\bibitem[\protect\citeauthoryear{{Lopez}, {Ellison}, {D'Odorico}  \&
  {Kim}}{{Lopez} et~al.}{2007}]{Lopez2007}
{Lopez} S.,  {Ellison} S.,  {D'Odorico} S.,   {Kim} T.-S.,  2007, \mn@doi
  [\aap] {10.1051/0004-6361:20065301}, \href
  {http://adsabs.harvard.edu/abs/2007A%26A...469...61L} {469, 61}

\bibitem[\protect\citeauthoryear{{Lopez} et~al.,}{{Lopez}
  et~al.}{2018}]{Lopez2018}
{Lopez} S.,  et~al., 2018, \mn@doi [\nat] {10.1038/nature25436}, \href
  {http://adsabs.harvard.edu/abs/2018Natur.554..493L} {554, 493}

\bibitem[\protect\citeauthoryear{{Lopez} et~al.,}{{Lopez}
  et~al.}{2020}]{Lopez2020}
{Lopez} S.,  et~al., 2020, \mn@doi [\mnras] {10.1093/mnras/stz3183}, \href
  {https://ui.adsabs.harvard.edu/abs/2020MNRAS.491.4442L} {491, 4442}

\bibitem[\protect\citeauthoryear{{Martin}, {Shapley}, {Coil}, {Kornei},
  {Bundy}, {Weiner}, {Noeske}  \& {Schiminovich}}{{Martin}
  et~al.}{2012}]{Martin2012}
{Martin} C.~L.,  {Shapley} A.~E.,  {Coil} A.~L.,  {Kornei} K.~A.,  {Bundy} K.,
  {Weiner} B.~J.,  {Noeske} K.~G.,   {Schiminovich} D.,  2012, \mn@doi [\apj]
  {10.1088/0004-637X/760/2/127}, \href
  {https://ui.adsabs.harvard.edu/abs/2012ApJ...760..127M} {760, 127}

\bibitem[\protect\citeauthoryear{{Martin}, {Ho}, {Kacprzak}  \&
  {Churchill}}{{Martin} et~al.}{2019}]{Martin2019}
{Martin} C.~L.,  {Ho} S.~H.,  {Kacprzak} G.~G.,   {Churchill} C.~W.,  2019,
  \mn@doi [\apj] {10.3847/1538-4357/ab18ac}, \href
  {https://ui.adsabs.harvard.edu/abs/2019ApJ...878...84M} {878, 84}

\bibitem[\protect\citeauthoryear{{Mitchell}, {Schaye}  \& {Bower}}{{Mitchell}
  et~al.}{2020}]{Mitchell2020}
{Mitchell} P.~D.,  {Schaye} J.,   {Bower} R.~G.,  2020, \mn@doi [\mnras]
  {10.1093/mnras/staa2252}, \href
  {https://ui.adsabs.harvard.edu/abs/2020MNRAS.497.4495M} {497, 4495}

\bibitem[\protect\citeauthoryear{{Mortensen}, {Keerthi Vasan}, {Jones},
  {Faucher-Gigu{\`e}re}, {Sanders}, {Ellis}, {Leethochawalit}  \&
  {Stark}}{{Mortensen} et~al.}{2021}]{Mortensen2021}
{Mortensen} K.,  {Keerthi Vasan} G.~C.,  {Jones} T.,  {Faucher-Gigu{\`e}re}
  C.-A.,  {Sanders} R.~L.,  {Ellis} R.~S.,  {Leethochawalit} N.,   {Stark}
  D.~P.,  2021, \mn@doi [\apj] {10.3847/1538-4357/abfa11}, \href
  {https://ui.adsabs.harvard.edu/abs/2021ApJ...914...92M} {914, 92}

\bibitem[\protect\citeauthoryear{{Nelson} et~al.,}{{Nelson}
  et~al.}{2020}]{Nelson2020}
{Nelson} D.,  et~al., 2020, \mn@doi [\mnras] {10.1093/mnras/staa2419}, \href
  {https://ui.adsabs.harvard.edu/abs/2020MNRAS.498.2391N} {498, 2391}

\bibitem[\protect\citeauthoryear{{Nielsen}, {Churchill}  \&
  {Kacprzak}}{{Nielsen} et~al.}{2013}]{Nielsen2013}
{Nielsen} N.~M.,  {Churchill} C.~W.,   {Kacprzak} G.~G.,  2013, \mn@doi [\apj]
  {10.1088/0004-637X/776/2/115}, \href
  {http://adsabs.harvard.edu/abs/2013ApJ...776..115N} {776, 115}

\bibitem[\protect\citeauthoryear{{Ohyama} et~al.,}{{Ohyama}
  et~al.}{2002}]{Ohyama2002}
{Ohyama} Y.,  et~al., 2002, \mn@doi [\pasj] {10.1093/pasj/54.6.891}, \href
  {https://ui.adsabs.harvard.edu/abs/2002PASJ...54..891O} {54, 891}

\bibitem[\protect\citeauthoryear{{P{\'e}roux} \& {Howk}}{{P{\'e}roux} \&
  {Howk}}{2020}]{Peroux2020}
{P{\'e}roux} C.,  {Howk} J.~C.,  2020, \mn@doi [\araa]
  {10.1146/annurev-astro-021820-120014}, \href
  {https://ui.adsabs.harvard.edu/abs/2020ARA&A..58..363P} {58, 363}

\bibitem[\protect\citeauthoryear{{P{\'e}roux}, {Rahmani}, {Arrigoni Battaia}
  \& {Augustin}}{{P{\'e}roux} et~al.}{2018}]{Peroux2018}
{P{\'e}roux} C.,  {Rahmani} H.,  {Arrigoni Battaia} F.,   {Augustin} R.,  2018,
  \mn@doi [\mnras] {10.1093/mnrasl/sly090}, \href
  {http://adsabs.harvard.edu/abs/2018MNRAS.479L..50P} {479, L50}

\bibitem[\protect\citeauthoryear{{Rahmani} et~al.,}{{Rahmani}
  et~al.}{2018}]{Rahmani2018}
{Rahmani} H.,  et~al., 2018, \mn@doi [\mnras] {10.1093/mnras/stx2726}, \href
  {http://adsabs.harvard.edu/abs/2018MNRAS.474..254R} {474, 254}

\bibitem[\protect\citeauthoryear{{Rauch}, {Sargent}, {Barlow}  \&
  {Carswell}}{{Rauch} et~al.}{2001}]{Rauch2001}
{Rauch} M.,  {Sargent} W.~L.~W.,  {Barlow} T.~A.,   {Carswell} R.~F.,  2001,
  \mn@doi [\apj] {10.1086/323523}, \href
  {http://adsabs.harvard.edu/abs/2001ApJ...562...76R} {562, 76}

\bibitem[\protect\citeauthoryear{{Rubin}, {Prochaska}, {Koo}, {Phillips},
  {Martin}  \& {Winstrom}}{{Rubin} et~al.}{2014}]{Rubin2014}
{Rubin} K. H.~R.,  {Prochaska} J.~X.,  {Koo} D.~C.,  {Phillips} A.~C.,
  {Martin} C.~L.,   {Winstrom} L.~O.,  2014, \mn@doi [\apj]
  {10.1088/0004-637X/794/2/156}, \href
  {https://ui.adsabs.harvard.edu/abs/2014ApJ...794..156R} {794, 156}

\bibitem[\protect\citeauthoryear{{Rubin}, {Diamond-Stanic}, {Coil}, {Crighton}
  \& {Moustakas}}{{Rubin} et~al.}{2018}]{Rubin2018a}
{Rubin} K.~H.~R.,  {Diamond-Stanic} A.~M.,  {Coil} A.~L.,  {Crighton} N.~H.~M.,
    {Moustakas} J.,  2018, \mn@doi [\apj] {10.3847/1538-4357/aa9792}, \href
  {http://adsabs.harvard.edu/abs/2018ApJ...853...95R} {853, 95}

\bibitem[\protect\citeauthoryear{{Rupke} et~al.,}{{Rupke}
  et~al.}{2019}]{Rupke2019}
{Rupke} D. S.~N.,  et~al., 2019, \mn@doi [\nat] {10.1038/s41586-019-1686-1},
  \href {https://ui.adsabs.harvard.edu/abs/2019Natur.574..643R} {574, 643}

\bibitem[\protect\citeauthoryear{{Schroetter} et~al.,}{{Schroetter}
  et~al.}{2019}]{Schroetter2019}
{Schroetter} I.,  et~al., 2019, \mn@doi [\mnras] {10.1093/mnras/stz2822}, \href
  {https://ui.adsabs.harvard.edu/abs/2019MNRAS.490.4368S} {490, 4368}

\bibitem[\protect\citeauthoryear{{Shaban} et~al.,}{{Shaban}
  et~al.}{2021}]{Shaban2021}
{Shaban} A.,  et~al., 2021, arXiv e-prints, \href
  {https://ui.adsabs.harvard.edu/abs/2021arXiv210913264S} {p. arXiv:2109.13264}

\bibitem[\protect\citeauthoryear{{Shopbell} \& {Bland-Hawthorn}}{{Shopbell} \&
  {Bland-Hawthorn}}{1998}]{Shopbell1998}
{Shopbell} P.~L.,  {Bland-Hawthorn} J.,  1998, \mn@doi [\apj] {10.1086/305108},
  \href {https://ui.adsabs.harvard.edu/abs/1998ApJ...493..129S} {493, 129}

\bibitem[\protect\citeauthoryear{{Soto}, {Lilly}, {Bacon}, {Richard}  \&
  {Conseil}}{{Soto} et~al.}{2016}]{Soto2016}
{Soto} K.~T.,  {Lilly} S.~J.,  {Bacon} R.,  {Richard} J.,   {Conseil} S.,
  2016, \mn@doi [\mnras] {10.1093/mnras/stw474}, \href
  {http://adsabs.harvard.edu/abs/2016MNRAS.458.3210S} {458, 3210}

\bibitem[\protect\citeauthoryear{{Steidel}, {Kollmeier}, {Shapley},
  {Churchill}, {Dickinson}  \& {Pettini}}{{Steidel} et~al.}{2002}]{Steidel2002}
{Steidel} C.~C.,  {Kollmeier} J.~A.,  {Shapley} A.~E.,  {Churchill} C.~W.,
  {Dickinson} M.,   {Pettini} M.,  2002, \mn@doi [\apj] {10.1086/339792}, \href
  {http://adsabs.harvard.edu/abs/2002ApJ...570..526S} {570, 526}

\bibitem[\protect\citeauthoryear{{Stewart}, {Brooks}, {Bullock}, {Maller},
  {Diemand}, {Wadsley}  \& {Moustakas}}{{Stewart} et~al.}{2013}]{Stewart2013}
{Stewart} K.~R.,  {Brooks} A.~M.,  {Bullock} J.~S.,  {Maller} A.~H.,  {Diemand}
  J.,  {Wadsley} J.,   {Moustakas} L.~A.,  2013, \mn@doi [\apj]
  {10.1088/0004-637X/769/1/74}, \href
  {https://ui.adsabs.harvard.edu/abs/2013ApJ...769...74S} {769, 74}

\bibitem[\protect\citeauthoryear{{Tejos} et~al.,}{{Tejos}
  et~al.}{2021}]{Tejos2021}
{Tejos} N.,  et~al., 2021, \mn@doi [\mnras] {10.1093/mnras/stab2147}, \href
  {https://ui.adsabs.harvard.edu/abs/2021MNRAS.507..663T} {507, 663}

\bibitem[\protect\citeauthoryear{{Tumlinson}, {Peeples}  \& {Werk}}{{Tumlinson}
  et~al.}{2017}]{Tumlinson2017}
{Tumlinson} J.,  {Peeples} M.~S.,   {Werk} J.~K.,  2017, \mn@doi [\araa]
  {10.1146/annurev-astro-091916-055240}, \href
  {http://adsabs.harvard.edu/abs/2017ARA%26A..55..389T} {55, 389}

\bibitem[\protect\citeauthoryear{{Weilbacher}, {Streicher}, {Urrutia}, {Jarno},
  {P{\'e}contal-Rousset}, {Bacon}  \& {B{\"o}hm}}{{Weilbacher}
  et~al.}{2012}]{Weilbacher2012}
{Weilbacher} P.~M.,  {Streicher} O.,  {Urrutia} T.,  {Jarno} A.,
  {P{\'e}contal-Rousset} A.,  {Bacon} R.,   {B{\"o}hm} P.,  2012, in
  {Radziwill} N.~M.,  {Chiozzi} G.,  eds,  Society of Photo-Optical
  Instrumentation Engineers (SPIE) Conference Series Vol. 8451, Software and
  Cyberinfrastructure for Astronomy II. p. 84510B, \mn@doi{10.1117/12.925114}

\bibitem[\protect\citeauthoryear{{Zabl} et~al.,}{{Zabl}
  et~al.}{2019}]{Zabl2019}
{Zabl} J.,  et~al., 2019, \mn@doi [\mnras] {10.1093/mnras/stz392}, \href
  {https://ui.adsabs.harvard.edu/abs/2019MNRAS.485.1961Z} {485, 1961}

\bibitem[\protect\citeauthoryear{{Zabl} et~al.,}{{Zabl}
  et~al.}{2020}]{Zabl2020}
{Zabl} J.,  et~al., 2020, \mn@doi [\mnras] {10.1093/mnras/stz3607}, \href
  {https://ui.adsabs.harvard.edu/abs/2020MNRAS.492.4576Z} {492, 4576}

\bibitem[\protect\citeauthoryear{{Zabl} et~al.,}{{Zabl}
  et~al.}{2021}]{Zabl2021}
{Zabl} J.,  et~al., 2021, \mn@doi [\mnras] {10.1093/mnras/stab2165}, \href
  {https://ui.adsabs.harvard.edu/abs/2021MNRAS.507.4294Z} {507, 4294}

\bibitem[\protect\citeauthoryear{{Zahedy}, {Chen}, {Rauch}, {Wilson}  \&
  {Zabludoff}}{{Zahedy} et~al.}{2016}]{Zahedy2016}
{Zahedy} F.~S.,  {Chen} H.-W.,  {Rauch} M.,  {Wilson} M.~L.,   {Zabludoff} A.,
  2016, \mn@doi [\mnras] {10.1093/mnras/stw484}, \href
  {http://adsabs.harvard.edu/abs/2016MNRAS.458.2423Z} {458, 2423}

\makeatother
\end{thebibliography}

\end{document}